\documentclass[aip,amsmath,amssymb,reprint,numerical]{revtex4-1}

\usepackage[utf8]{inputenc}
\usepackage[T1]{fontenc}
\usepackage{mathptmx}         
\usepackage{etoolbox}
\usepackage{physics}
\usepackage{braket}
\usepackage{tikz}
\usetikzlibrary{arrows}
\usepackage{graphicx}
\usepackage{float}
\usepackage{xcolor}

\usepackage{silence}
\usepackage{hyperref}         

\newcommand{\maybeincludegraphics}[2][]{%
\IfFileExists{#2}{\includegraphics[#1]{#2}}{%
\fbox{\parbox{0.86\columnwidth}{\centering Missing figure file:\\ \texttt{\detokenize{#2}}}}%
}}

\makeatletter
\def\@email#1#2{%
 \endgroup
 \patchcmd{\titleblock@produce}
  {\frontmatter@RRAPformat}
  {\frontmatter@RRAPformat{\produce@RRAP{*#1\href{mailto:#2}{#2}}}\frontmatter@RRAPformat}
  {}{}
}%
\makeatother
\begin{document}

\preprint{AIP/123-QED}

\title{Kubo-Martin-Schwinger-Gated Imaginary-Time Reconstruction of Time-Resolved Electronic Circular Dichroism in Organic Excitonic Aggregates}

\author{Christian Tantardini*}
\affiliation{Department of Materials Science and NanoEngineering, Rice University, Houston, Texas 77005, United States of America.}
\affiliation{Center for Integrative Petroleum Research, King Fahd University of Petroleum and Minerals, Dhahran 31261, Saudi Arabia}

\author{Caterina Cocchi*}
\affiliation{Friedrich-Schiller-Universit\"at Jena, Institut für Festk\"orpertheorie und -Optik,  07743 Jena, Germany}

\email{christiantantardini@ymail.com; caterina.cocchi@uni-jena.de}

\date{\today}

\begin{abstract}
Time-resolved electronic circular dichroism (TR-ECD) and time-resolved
circular dichroism (TRCD) probe chiral exciton dynamics prepared by an
ultrafast pump and interrogated by a weak circular probe. The measured signal
is a causal mixed electric--magnetic response, whereas imaginary-time methods
require a sufficiently stationary reference. Here, we introduce a
Kubo--Martin--Schwinger-gated criterion that determines, at each pump--probe
delay, whether the selected TRCD-like response can be represented by a
conditional one-exciton Gibbs ensemble or requires explicit real-time or
Keldysh dynamics. The diagnostic combines a state-level distance from the
conditional reference with an integrated spectral distance for the reciprocal
E1--M1 response. We apply the protocol to three literature-constrained
Frenkel-exciton models: a Ress-type squaraine-polymer squeezed helix, a cisoid
indolenine squaraine B hexamer, and a helical perylene-bisimide stack. The
results show that early nonthermal states are non-admissible, whereas
intramanifold population relaxation and the decay of coherence memory can
open a Matsubara-admissible delay window before complete excited-state
recovery. The resulting gate identifies when imaginary-time methods can be
used without imposing a stationary description on a genuinely nonequilibrium
chiroptical response.
\end{abstract}

\maketitle

\section{Introduction}
\label{sec:introduction}

Time-resolved circular dichroism (TRCD) is a powerful ultrafast pump–probe spectroscopic technique that tracks chiral excited-state dynamics and conformational changes with femtosecond time resolution \cite{Stadnytskyi2017_TRCD_spectrometer,Tapavicza2023_AbInitio_TRCD,Moffitt1995_TimeResolvedECD}.
Early work demonstrated TRCD in pigment complexes, revealing excitonic dynamics inaccessible to standard transient absorption \cite{Savikhin2000_TRCD_Pigments}.
TRCD has also been applied in protein folding studies \cite{Gifford1999_TRCDProteins}, and theoretical frameworks have linked chiral electronic structure evolution to transient CD spectra \cite{Cho2003_2DChiroptical,Rouxel2017_Xray_TRCD}.
Extensions to spin and semiconductor systems further illustrate the breadth of time-resolved circular dichroism as a dynamic chiral probe \cite{Aktsipetrov2008_TRCD_Semiconductors}. 

In molecular aggregates, the chiroptical response is intrinsically linked to the underlying excitonic structure, which arises from chromophore alignment, intersite coupling, and collective transition dipoles typical of H- and J-aggregate architectures \cite{Wurthner2011_Angew, Hestand2018_ChemRev, Spano2010_ACR, Deshmukh2022_ChemPhysRev, Tempelaar2017_JPhysChemLett}.
The degree of exciton delocalization, the presence of static disorder, and the spatial arrangement of transition dipoles strongly shape both linear and chiroptical spectra, as demonstrated in theoretical studies of one-dimensional and helical aggregates \cite{Didraga2004_JPhysChemB, YuenZhou2014_JChemPhys, Ivanov2010_ChemPhysLett}.
For example, perylene bisimide aggregates exhibit size-dependent exciton dynamics that manifest in their optical and chiroptical signatures \cite{Wolter2012_JPCC, Schroter2013_JChemPhys}, and engineered exciton circuits in synthetic architectures highlight controlled energy propagation relevant for chiral optical responses \cite{Banal2017_JPhysChemLett}.
TRCD and related ultrafast chiral spectroscopies have proven capable of tracking these coherent excitonic features on femto- to picosecond timescales, revealing dynamics that are often inaccessible in standard transient absorption spectroscopy \cite{Savikhin2000_TRCD_Pigments, Stadnytskyi2017_TRCD_spectrometer}.
In particular, recent TRCD studies on cisoid squaraine hexamers and helical perylene-bisimide stacks have directly observed exciton localization and ultrafast coherent motion, underscoring the need for robust theoretical frameworks that bridge real-time exciton dynamics with many-body electronic structure and response theory \cite{Fischermeier2024_PCCP, Sung2015_NatCommun}.

The main theoretical challenge is that, while a weak circular probe measures a \textit{causal} retarded response of a pump-prepared ensemble, most of the existing electronic-structure many-body methods, such as Green-function theory~\cite{schindlmayr2000decay}, quantum Monte Carlo~\cite{baroni1999reptation}, dynamical mean-field theory~\cite{wolf2015imaginary}, numerical renormalization group~\cite{bulla2001finite}, and Matsubara-frequency approaches~\cite{shinaoka2017compressing,kaye2022discrete}, are formulated in imaginary time. These methods are ideally suited for stationary states, but are not directly applicable to the nonthermal, coherently driven distributions typical of the early-time pump–probe regime. Reconstructing a real-time signal from imaginary-time data requires analytic continuation. This is an intrinsically ill-conditioned process that demands a well-defined stationary reference satisfying the Kubo–Martin–Schwinger (KMS) condition \cite{JarrellGubernatis1996_PhysRep}. 


The central question addressed in this work is when a pump-prepared exciton
ensemble has relaxed sufficiently for an imaginary-time description to become
admissible. Immediately after photoexcitation, nonthermal populations and
residual coherences retain information about the pump preparation. Replacing
this state by a stationary Matsubara reference would erase these
nonequilibrium signatures. At later delays, dephasing and intramanifold
relaxation can drive the optically active one-exciton block toward a
quasi-stationary distribution even though the full system has not returned to
global equilibrium.

We therefore formulate a KMS-gated reconstruction protocol for TRCD in
molecular aggregates described by Frenkel excitons. The protocol combines a
state-level comparison with a conditional one-exciton Gibbs reference and an
observable-level validation based on the selected reciprocal E1--M1 response.
It is not intended as a universal measure of nonequilibrium. Rather, it is an
operational criterion for deciding whether a given pump--probe delay can be
treated using an imaginary-time reference or requires explicit real-time
dynamics.

We test the protocol on three literature-constrained benchmarks: (i) a Ress-type squaraine-polymer squeezed helix \cite{Ress2023_ChemSci}, (ii) a cisoid indolenine squaraine B hexamer  \cite{Fischermeier2024_PCCP}, and  a helical perylene-bisimide stack exhibiting coherent exciton dynamics \cite{Sung2015_NatCommun}. 
Our results demonstrate that admissibility is not a static property of the aggregate but a dynamic variable dependent on the pump-probe delay and the specific response channel. 
By identifying these ``admissibility windows'', our workflow provides a practical
gate for utilizing efficient imaginary-time approaches in chiroptical
spectroscopy without compromising essential nonequilibrium information. The
resulting protocol should be understood as a two-level gate:
\(\Delta_{\rm state}^{X}\) provides an a priori state-based screening criterion,
whereas \(D_{\rm spec}^{X}\) provides an observable-level validation of that
screening for the selected TRCD-like response.

\section{Theory and reconstruction criterion}
\label{sec:theory}

The framework has three elements. First, the TRCD signal is identified as the
retarded mixed electric--magnetic response of the pump-prepared ensemble.
Second, the stationarity requirement underlying imaginary-time reconstruction
is formulated through the KMS condition. Third, a conditional-exciton
diagnostic is introduced to identify the pump--probe delays at which a
stationary one-exciton reference becomes admissible for the selected response
channel.

\subsection{Weak-probe TR-ECD/TRCD response}
\label{sec:weak_probe_response}

To determine the validity of imaginary-time methods in the transient regime, we distinguish between a formal many-body KMS condition and a practical conditional-exciton admissibility criterion.
We consider a pump--probe experiment in which a circularly polarized pump
prepares a nonequilibrium excited state and a weak circular probe interrogates
the subsequent chiroptical response. During the preparation stage, the
Hamiltonian is written as
\begin{equation}
H(t)=H_0+H_{\rm pump}(t),
\end{equation}
with the pump interaction retained to electric-dipole--magnetic-dipole
(E1--M1) order,
\begin{equation}
H_{\rm pump}(t)
=
-f_{\rm p}(t-t_{\rm p})
\left[
\hat{\boldsymbol{\mu}}\cdot\mathbf E_{\rm p}
+
\hat{\mathbf m}\cdot\mathbf B_{\rm p}
\right].
\label{eq:pump_interaction}
\end{equation}
Here, \(f_{\rm p}(t-t_{\rm p})\) is the pump envelope, \(t_{\rm p}\) is its
maximum or envelope centroid, and \(\mathbf E_{\rm p}\) and
\(\mathbf B_{\rm p}\) are the electric and magnetic components of the
circularly polarized pump. The pump--probe delay is
\begin{equation}
\tau=t_0-t_{\rm p},
\end{equation}
where \(t_0\) is the center of the probe pulse.

Immediately before the probe arrives, the density matrix is:
\begin{align}
\rho_\tau
&=
U(t_0,t_{\rm ini})\,\rho_{\rm ini}\,U^\dagger(t_0,t_{\rm ini}),
\\
U(t_0,t_{\rm ini})
&=
\mathcal{T}
\exp\!\left[
-\frac{i}{\hbar}
\int_{t_{\rm ini}}^{t_0}H(t)\,dt
\right].
\label{eq:rho_tau}
\end{align}

Here, \(t_{\rm ini}\) is a time before the pump field is applied, so that
\(\rho_{\rm ini}\) denotes the pre-pump equilibrium or otherwise prepared
initial ensemble.

The probe is assumed weak enough that its action can be treated perturbatively
to first order. For a circularly polarized probe, the E1--M1 interaction can be
written as
\begin{equation}
H_{\rm pr}(t)
=
-
g(t-t_0)
\left[
\hat{\boldsymbol{\mu}}\cdot\mathbf E_0
+
\hat{\mathbf m}\cdot\mathbf B_0
\right],
\label{eq:probe_perturbation}
\end{equation}
where \(g(t)\) is the probe envelope, \(\mathbf E_0\) and \(\mathbf B_0\) are
the electric and magnetic field amplitudes of the probe, and
\(\hat{\boldsymbol{\mu}}\) and \(\hat{\mathbf m}\) are the electric- and
magnetic-dipole operators. In the ideal impulsive limit,
\(g(t)\rightarrow\delta(t)\)~\cite{guandalini2021nonlinear}. For a finite probe
pulse, \(g(t)\) is retained explicitly through its transfer function.

Linear-response theory yields the induced magnetic and electric responses after
the probe as mixed E1--M1 response functions. For example, the magnetic response
induced by the electric part of the probe is
\begin{equation}
m_k^{\rm ind}(t)
=
\int_{-\infty}^{\infty}
dt'\,
\chi^R_{m_k\mu_j}(t-t';\tau)\,
E_{0,j} g(t'),
\label{eq:weak_probe_convolution}
\end{equation}
where the retarded mixed electric--magnetic susceptibility of the pump-prepared
ensemble is
\begin{equation}
\chi^R_{m_k\mu_j}(t;\tau)
=
\frac{i}{\hbar}\,
\theta(t)\,
{\rm Tr}
\left[
\rho_\tau
\left[
\hat{m}_k(t),\hat{\mu}_j(0)
\right]
\right].
\label{eq:retarded_mumu}
\end{equation}
The reciprocal electric response induced by the magnetic part of the probe is
described by the corresponding susceptibility \(\chi^R_{\mu_k m_j}\).

Here, the operators are evolved with the post-pump Hamiltonian used during the probe window. Eq.~\eqref{eq:retarded_mumu} is the central relation of this work, which must be satisfied by the imaginary-time reconstruction. It is worth stressing that reproducing this kernel is a more stringent requirement than matching time-averaged experimental spectra, as it preserves the phase and causality information of the excitation.

Fourier-transforming Eq.~\eqref{eq:weak_probe_convolution} gives
\begin{equation}
\widetilde{m}_k^{\rm ind}(\omega)
=
E_{0,j}\,
\widetilde{g}(\omega)\,
\chi^R_{m_k\mu_j}(\omega;\tau),
\label{eq:probe_transfer}
\end{equation}

with
\begin{equation}
\chi^R_{m_k\mu_j}(\omega;\tau)
=
\int_{0}^{\infty}
dt\,
e^{i\omega t}
\chi^R_{m_k\mu_j}(t;\tau).
\label{eq:retarded_frequency}
\end{equation}
Thus, within the weak-probe regime, a finite probe pulse multiplies the material response by a known transfer function without changing the definition of the retarded kernel. Strong probe fields, pump--probe temporal overlap, or nonlinear probe effects lie outside this factorized limit.

For an isotropic ensemble, the TR-ECD/TRCD signal is controlled by the
orientationally averaged E1--M1 pseudoscalar response. In compact form,
\begin{equation}
I_{\rm TR\mbox{-}ECD}(\omega;\tau)
\propto
\frac{\omega}{c}
\,{\rm Im}\,
\frac{1}{3}
\sum_k
\left[
\chi^{R}_{m_k\mu_k}(\omega;\tau)
+
\chi^{R}_{\mu_k m_k}(\omega;\tau)
\right],
\label{eq:continuum_trecd_signal}
\end{equation}
where the two mixed susceptibilities represent the reciprocal electric--magnetic
linear-response channels generated by the weak circular probe.

In the reduced Frenkel-exciton implementation, the mixed E1--M1 strength of
exciton eigenstate \(\alpha\) is constructed, up to a common normalization,
from the origin-independent pairwise exciton-chirality expression
\begin{equation}
\mathcal R_{\alpha}
=
\frac{E_{\alpha}}{2\hbar c}
\sum_{m,n}
C_{m\alpha}C_{n\alpha}
\left(\mathbf r_m-\mathbf r_n\right)
\cdot
\left(
\boldsymbol{\mu}_m\times\boldsymbol{\mu}_n
\right).
\label{eq:reduced_mixed_strength}
\end{equation}
The reduced benchmark model is reciprocal, and the total mixed strength is
therefore divided equally between the two response channels,
\begin{equation}
\mathcal R_{\alpha}^{m\mu}
=
\mathcal R_{\alpha}^{\mu m}
=
\frac{1}{2}\mathcal R_{\alpha}.
\label{eq:reciprocal_channel_strengths}
\end{equation}
The numerical TRCD-like spectrum is evaluated as the sum
\begin{equation}
I_{\rm TRCD}(\omega;\tau)
=
I_{m\mu}(\omega;\tau)
+
I_{\mu m}(\omega;\tau),
\label{eq:reduced_trcd_sum}
\end{equation}
so that both reciprocal terms appearing in
Eq.~\eqref{eq:continuum_trecd_signal} are retained explicitly. The two
channel-resolved spectra are also written separately to the numerical output
files. Their equality is a property of the present reciprocal reduced model
and is not assumed to hold for a general tensorial or \textit{ab initio}
implementation.

The magnetic dipole can be expressed from the physical current density as:
\begin{equation}
\hat{m}_k
=
\frac{1}{2}
\int d^3r\,
\varepsilon_{k\ell m}
r_\ell
\hat{j}_m(\mathbf r).
\label{eq:magnetic_from_current}
\end{equation}
While Eq.~\eqref{eq:magnetic_from_current} represents the general microscopic definition accessible from first principles, when dealing with molecular aggregates, it is more convenient to utilize a projected representation, retaining the same response structure by introducing effective electric and magnetic transition operators:
\begin{align}
\hat{\mu}
&=
\sum_i
\left(
\mu_i \ket{i}\bra{g}
+
\mu_i^* \ket{g}\bra{i}
\right),
\\
\hat{m}
&=
\sum_i
\left(
m_i \ket{i}\bra{g}
+
m_i^* \ket{g}\bra{i}
\right),
\label{eq:effective_mu_m}
\end{align}

where \(\ket{g}\) is the electronic ground state and \(\ket{i}\) is a
Frenkel-exciton state. In the benchmark results, \(\hat m\) is an effective
magnetic-transition operator that generates the E1--M1 chiroptical channel.

The magnetic-dipole operator is known to raise gauge-origin issues in molecular
ECD calculations when used in a truncated basis. In the present benchmark
calculations, \(\hat m\) is not used as an ab initio current-density magnetic
moment. It is an effective magnetic-transition operator defining a reduced
E1--M1 chiroptical response channel. The same operator and origin convention are
used for the pump-prepared and conditional-reference spectra; therefore, the
admissibility comparison is internally origin-consistent. A fully ab initio
implementation of the present protocol should use a gauge-origin-invariant
formulation, for example through current-density response or equivalent
gauge-including approaches.
In reduced response calculations, Eq.~\eqref{eq:effective_mu_m} provides the mixed operator pair whose retarded response is compared between the pump-prepared ensemble and the conditional stationary reference.
The key consequence is common to both the continuum and reduced exciton descriptions: TR-ECD/TRCD is a retarded response measurement. Any imaginary-time treatment must therefore reproduce the causal retarded kernel after analytic continuation, and not merely fit the appearance of a finite-window spectrum. In the presence of strong pump--probe overlap or non-perturbative fields, the distinction between material response and pulse envelope blurs, necessitating a full Keldysh treatment.

\subsection{Imaginary-time reconstruction and conditional-exciton admissibility}
\label{sec:kms_diagnostics}

The formal route from imaginary time to real frequency is standard when the
state is stationary and satisfies the KMS condition. To keep the frequency
dimensions explicit, we introduce the bosonic Matsubara energies
\(\nu_n=2\pi n/\beta\). For a stationary reference ensemble,
\(\rho_{\rm ref}\propto\exp(-\beta H_{\rm ref})\), the mixed Matsubara
correlator is
\begin{equation}
\chi_{m_k\mu_j}(i\nu_n)
=
\int_0^\beta
d\lambda\,
e^{i\nu_n\lambda}
\left\langle
T_\lambda
\hat m_k(\lambda)
\hat\mu_j(0)
\right\rangle_{\rho_{\rm ref}},
\label{eq:matsubara_mixed}
\end{equation}
where
\(\hat m_k(\lambda)
=
e^{\lambda H_{\rm ref}}
\hat m_k
e^{-\lambda H_{\rm ref}}\).
Analytic continuation to the real-frequency response is then
\begin{equation}
\chi^R_{m_k\mu_j}(\omega)
=
\chi_{m_k\mu_j}
\!\left(
i\nu_n\rightarrow\hbar\omega+i0^+
\right).
\label{eq:matsubara_to_retarded}
\end{equation}
In a pump--probe experiment, however, \(\rho_\tau\) is generally
nonequilibrium and need not satisfy the stationarity requirements of a Gibbs
ensemble. Imaginary-time reconstruction must therefore be gated by a
diagnostic that tests whether a stationary reference is admissible at the
delay under consideration.

For the general response problem, let
\(S^>_{m\mu}(\omega;\tau)\) and
\(S^<_{m\mu}(\omega;\tau)\) denote the Fourier transforms of
\(\langle \hat m(t)\hat\mu(0)\rangle_{\rho_\tau}\) and
\(\langle \hat\mu(0)\hat m(t)\rangle_{\rho_\tau}\), respectively, using the
same Fourier convention as Eq.~\eqref{eq:retarded_frequency}. With this
convention, a stationary KMS state satisfies
\(S^>_{m\mu}(\omega;\tau)
=
e^{\beta\hbar\omega}
S^<_{m\mu}(\omega;\tau)\).
The reciprocal \(\mu m\) channel obeys the analogous relation. A complete KMS
certification of the circular response must therefore test both reciprocal
mixed channels. For compactness, the following defect is written explicitly
for the \(m\mu\) channel:
\begin{equation}
\Delta_{\rm KMS}(\tau)
=
\frac{
\int d\omega\,
w(\omega)
\left|
S^>_{m\mu}(\omega;\tau)
-
e^{\beta\hbar\omega}
S^<_{m\mu}(\omega;\tau)
\right|
}{
\int d\omega\,
w(\omega)
\left|
S^>_{m\mu}(\omega;\tau)
\right|
+\epsilon
},
\label{eq:delta_kms_general}
\end{equation}
where \(w(\omega)\) selects the spectral window of interest and \(\epsilon\)
is a positive regularization constant. A small defect indicates that the
state is sufficiently stationary within that window to admit a Matsubara
reconstruction; a large defect identifies a regime requiring explicit
real-time propagation or a Keldysh-contour treatment.

In molecular aggregates, the optical gap is typically larger than
\(1~{\rm eV}\) and therefore much greater than \(k_{\rm B}T\) at room
temperature. The equilibrium Gibbs ensemble consequently contains a
negligible one-exciton population. A global equilibrium KMS test would then
characterize predominantly the unexcited system rather than the
pump-populated exciton manifold probed in a transient experiment. The
physically relevant question is instead whether the normalized one-exciton
block has reached a quasi-stationary conditional distribution before
radiative or nonradiative recombination removes the transient population.
The organic benchmarks therefore use the conditional-exciton diagnostic
defined in Eqs.~\eqref{eq:delta_pop_x}--\eqref{eq:delta_adm_x}, rather than
the full-space defect of Eq.~\eqref{eq:delta_kms_general}, as the plotted
admissibility measure.

Let \(\hat{\Pi}_X\) denote the projector onto the one-exciton manifold.
The normalized excited-state block and the corresponding one-exciton
population are
\begin{align}
\rho_X(\tau)
&=
\frac{
\hat{\Pi}_X\rho_\tau\hat{\Pi}_X
}{
{\rm Tr}\!\left(
\hat{\Pi}_X\rho_\tau\hat{\Pi}_X
\right)
},
\\
P_X(\tau)
&=
{\rm Tr}\!\left(
\hat{\Pi}_X\rho_\tau
\right).
\label{eq:normalized_exciton_block}
\end{align}
Let \(H_X\) be the Frenkel Hamiltonian restricted to this manifold. The
conditional Gibbs reference used in the benchmark calculations is
\begin{equation}
\rho_X^{\rm G}
=
\frac{
\exp(-\beta H_X)
}{
{\rm Tr}_X\!\left[
\exp(-\beta H_X)
\right]
},
\qquad
\beta=\frac{1}{k_{\rm B}T},
\label{eq:conditional_gibbs_ref}
\end{equation}
where \({\rm Tr}_X\) denotes the trace over the one-exciton states. In the
present calculations, the benchmark bath temperature is fixed at
\(T=300~{\rm K}\) for all three aggregates.

In the exciton eigenbasis, we denote the diagonal populations obtained from
Eq.~\eqref{eq:conditional_gibbs_ref} by
\(p_\alpha^{\rm G}
=
\langle\alpha|\rho_X^{\rm G}|\alpha\rangle\).
The conditional-reference spectrum at delay \(\tau\) is evaluated with
\begin{equation}
P_\alpha^{\rm G}(\tau)
=
P_X(\tau)\,p_\alpha^{\rm G}.
\label{eq:conditional_reference_population}
\end{equation}
The pump-prepared and conditional-reference spectra therefore contain the
same total one-exciton population and differ only in their normalized
intramanifold distributions.

\begin{figure*}[!htbp]
  \centering
  \maybeincludegraphics[width=1.0\textwidth]{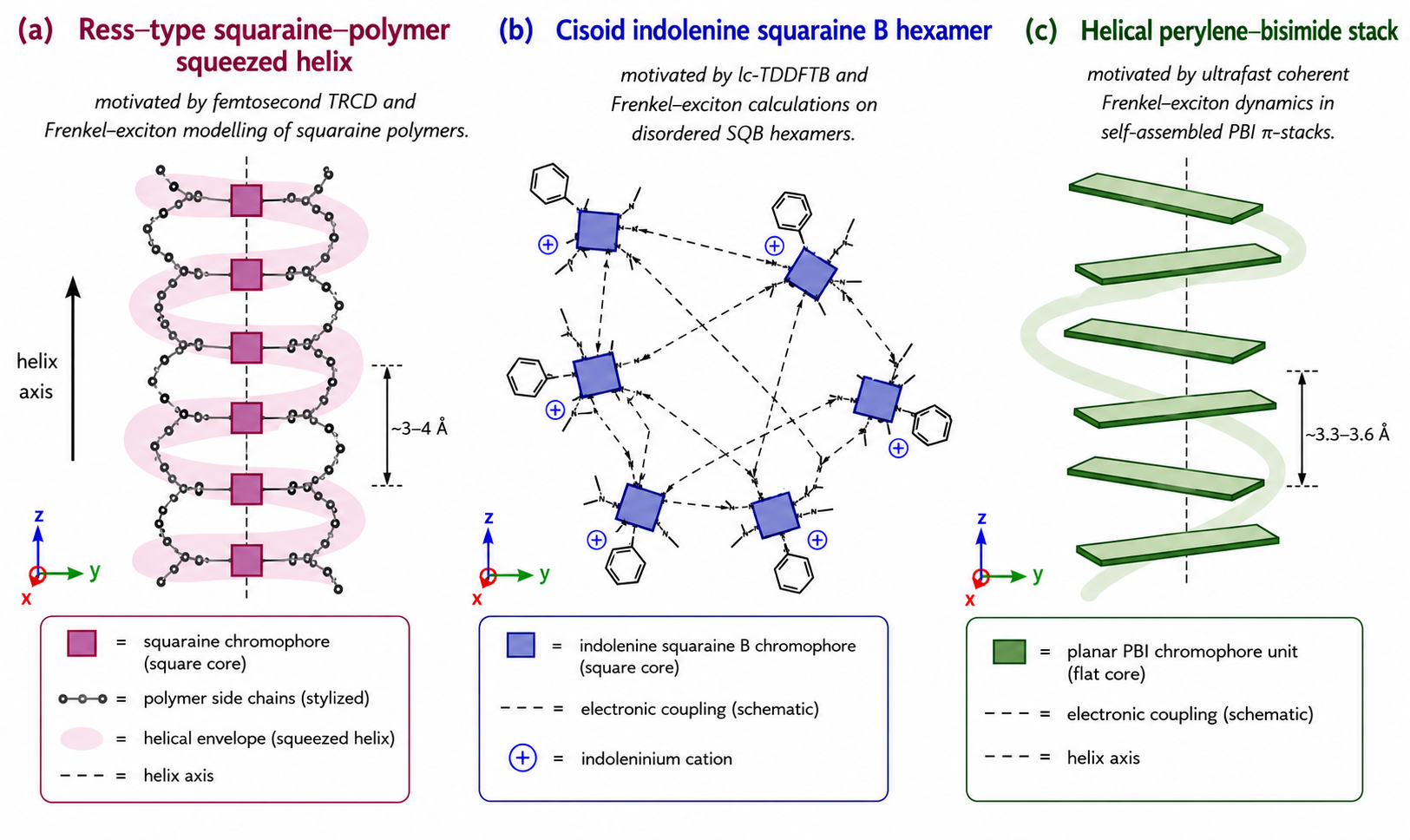}
  \caption{
  Schematic overview of the organic Frenkel-exciton benchmark classes used in this work.
  (a) Ress squaraine-polymer squeezed helix, motivated by femtosecond TRCD and Frenkel-exciton modelling of squaraine polymers.
  (b) Cisoid indolenine squaraine B hexamer, motivated by lc-TDDFTB and Frenkel-exciton calculations on disordered SQB hexamers.
  (c) Helical perylene-bisimide stack, motivated by ultrafast coherent Frenkel-exciton dynamics in self-assembled PBI \(\pi\)-stacks.
  The drawings are qualitative cartoons of chromophore organization and excitonic coupling topology, not atomistic structural models.
  }
  \label{fig:organic_benchmark_overview}
\end{figure*}

The state-level distance from this conditional reference is measured by the
exciton population defect
\begin{equation}
\Delta_{\rm pop}^{X}(\tau)
=
\frac{1}{2}
\sum_\alpha
\left|
p_\alpha(\tau)
-
p_\alpha^{\rm G}
\right|,
\label{eq:delta_pop_x}
\end{equation}
where
\(p_\alpha(\tau)
=
\langle\alpha|\rho_X(\tau)|\alpha\rangle\).
The residual coherence is measured from the off-diagonal part of the
normalized excited-state density matrix,
\begin{equation}
\Delta_{\rm coh}^{X}(\tau)
=
\frac{
\left\|
\rho_X(\tau)-{\rm diag}\rho_X(\tau)
\right\|_{\rm F}
}{
\left\|
\rho_X(\tau)
\right\|_{\rm F}
+\epsilon
},
\label{eq:delta_coh_x}
\end{equation}
and the state admissibility defect is:
\begin{equation}
\Delta_{\rm state}^{X}(\tau)
=
\min\!\left\{
1,\,
\left[
\left(\Delta_{\rm pop}^{X}(\tau)\right)^2
+
\left(\gamma\,\Delta_{\rm coh}^{X}(\tau)\right)^2
\right]^{1/2}
\right\}.
\label{eq:delta_state_x}
\end{equation}
In Eq.~\eqref{eq:delta_state_x}, \(\gamma\) controls the relative weight of
the population and coherence contributions.

For the reduced benchmark calculations, the full off-diagonal density matrix
is not propagated explicitly. Consequently,
Eq.~\eqref{eq:delta_coh_x} is not evaluated directly. Instead, the state-level
gate uses the phenomenological coherence-memory proxy
\begin{equation}
\mathcal C_X(\tau)
=
\min\!\left[
1,\,
\left(
\sum_\alpha
\widetilde p_\alpha^{\,2}(0)
\right)^{1/2}
\exp\!\left(
-\frac{\tau}{T_2}
\right)
\right],
\label{eq:coherence_proxy_benchmarks}
\end{equation}
where
\(\widetilde p_\alpha(0)=P_\alpha(0)/P_X(0)\) and \(T_2\) is the benchmark
coherence-decay time. The prefactor is the square root of the initial
population purity and scales the phenomenological memory amplitude; it is not
an independently propagated Frobenius norm. In the benchmark evaluation of
Eq.~\eqref{eq:delta_state_x}, we therefore set
\(\Delta_{\rm coh}^{X}(\tau)\rightarrow\mathcal C_X(\tau)\) and
\(\gamma=0.75\).

The observable-level defect is defined directly from the benchmark spectra. With \(I_{\rm pp}(\omega;\tau)\) denoting the pump-prepared retarded TRCD-like spectrum and \(I_{\rm G}(\omega;\tau)\) the corresponding spectrum computed from the conditional Gibbs reference, the normalized spectral distance becomes:
\begin{equation}
D_{\rm spec}^{X}(\tau)
=
\frac{
\int d\omega\,
\left|
I_{\rm pp}(\omega;\tau)-I_{\rm G}(\omega;\tau)
\right|
}{
\int d\omega\,
\left|
I_{\rm pp}(\omega;\tau)
\right|
+\epsilon
}.
\label{eq:d_spec_x}
\end{equation}

In the numerical implementation, Eq.~\eqref{eq:d_spec_x} is evaluated as an
integrated absolute \(L^1\) distance over the common probe-energy interval
\(1.55\leq E\leq2.80~{\rm eV}\), using the same energy grid for the
pump-prepared and conditional-reference spectra. Writing the integral in
terms of \(E=\hbar\omega\) does not change the normalized ratio because the
common Jacobian cancels between numerator and denominator.

The benchmark calculations do not solve an inverse analytic-continuation
problem. They test the admissibility criterion by comparing two retarded
spectra generated with the same reduced E1--M1 operators and broadening model,
but with the pump-prepared and conditional Gibbs populations, respectively.
The resulting \(D_{\rm spec}^{X}\) therefore validates the state-based gate;
it does not assess the numerical stability of a particular continuation
algorithm.

The final benchmark diagnostic is the conservative envelope:
\begin{equation}
\Delta_{\rm adm}^{X}(\tau)
=
\min\!\left\{
1,\,
\max\!\left[
\Delta_{\rm state}^{X}(\tau),
D_{\rm spec}^{X}(\tau)
\right]
\right\}.
\label{eq:delta_adm_x}
\end{equation}

The quantities entering Eq.~\eqref{eq:delta_adm_x} play different practical
roles. The state-level defect \(\Delta_{\rm state}^{X}\) is an a priori
screening quantity: it can be evaluated from the pump-prepared one-exciton
density matrix, the Frenkel Hamiltonian, and the conditional Gibbs reference defined at the benchmark bath temperature. It does not require computing the
full TRCD-like spectrum with both approaches. In contrast, \(D_{\rm spec}^{X}\)
is an observable-level validation quantity. It is used in the benchmarks to test
whether the selected E1--M1 response channel is actually reproduced by the
conditional stationary reference. The practical workflow is therefore
hierarchical. A delay is first screened using \(\Delta_{\rm state}^{X}\); only
after this state-level gate is passed is the imaginary-time reference expected
to be reliable. In benchmark calculations, or in applications requiring
quantitative certification, \(D_{\rm spec}^{X}\) can then be evaluated as an
a posteriori validation of the selected observable.

This criterion is fundamentally different from the KMS condition introduced in
Eq.~\eqref{eq:delta_kms_general}: \(\Delta_{\rm KMS}\) tests full detailed
balance of greater and lesser spectra, whereas \(\Delta_{\rm adm}^{X}\) tests
whether the conditional one-exciton reference is sufficient to reconstruct the
specific TRCD-like response of interest.

In the present benchmarks, \(\Delta_{\rm state}^{X}\) is usually the limiting
contribution to the envelope. This need not be true in general. The spectral
defect can become limiting when the selected TRCD-like observable is controlled
by weak transitions, near-cancellation between positive and negative chiroptical
bands, vibronic sidebands, narrow spectral windows, or magnetic-transition
contributions that are more sensitive to residual density-matrix structure than
the population norm itself. In such cases, a state that appears close to the
conditional Gibbs reference may still yield a measurable chiroptical line-shape
error.

Causality remains a separate requirement for any retarded response obtained
by analytic continuation. In a full imaginary-time implementation, the
continued spectrum should be checked against the relevant commutator sum
rules and Kramers--Kronig relations. These quantities are not used as
independent numerical diagnostics in the present reduced benchmarks, whose
purpose is to test the delay-dependent admissibility gate defined in
Eq.~\eqref{eq:delta_adm_x}.

\section{Organic Frenkel-exciton benchmark construction}
\label{sec:benchmark_construction}

We evaluate the conditional-exciton admissibility criterion using three
literature-constrained Frenkel-exciton models. These systems are not intended
as complete reproductions of the corresponding experiments; they provide
physically motivated test cases for resolving the delay-dependent transition
from a pump-prepared nonequilibrium distribution to a conditional stationary
one-exciton ensemble.
The three benchmark systems are visualized in Fig.~\ref{fig:organic_benchmark_overview}. Benchmark A is a Ress-type squaraine-polymer squeezed helix, motivated by the femtosecond TRCD/Frenkel-exciton study of Ress \emph{et al.}, where the TRCD signal is defined as the difference between transient absorption probed with left- and right-circularly polarized light and where exciton relaxation and slower ground-state recovery occur on separated time scales \cite{Ress2023_ChemSci}. Benchmark B is a cisoid indolenine squaraine B hexamer, motivated by the lc-TDDFTB and Frenkel-exciton study of Fischermeier \emph{et al.}, where molecular-dynamics-generated disordered structures were used to analyse absorption and exciton localization in SQB hexamers \cite{Fischermeier2024_PCCP}. Benchmark C is a helical perylene-bisimide stack, motivated by the direct observation of ultrafast coherent Frenkel-exciton dynamics in self-assembled PBI \(\pi\)-stacks by Sung \emph{et al.} \cite{Sung2015_NatCommun}.

\begin{figure}[!t]
  \centering
  \maybeincludegraphics[width=\columnwidth]{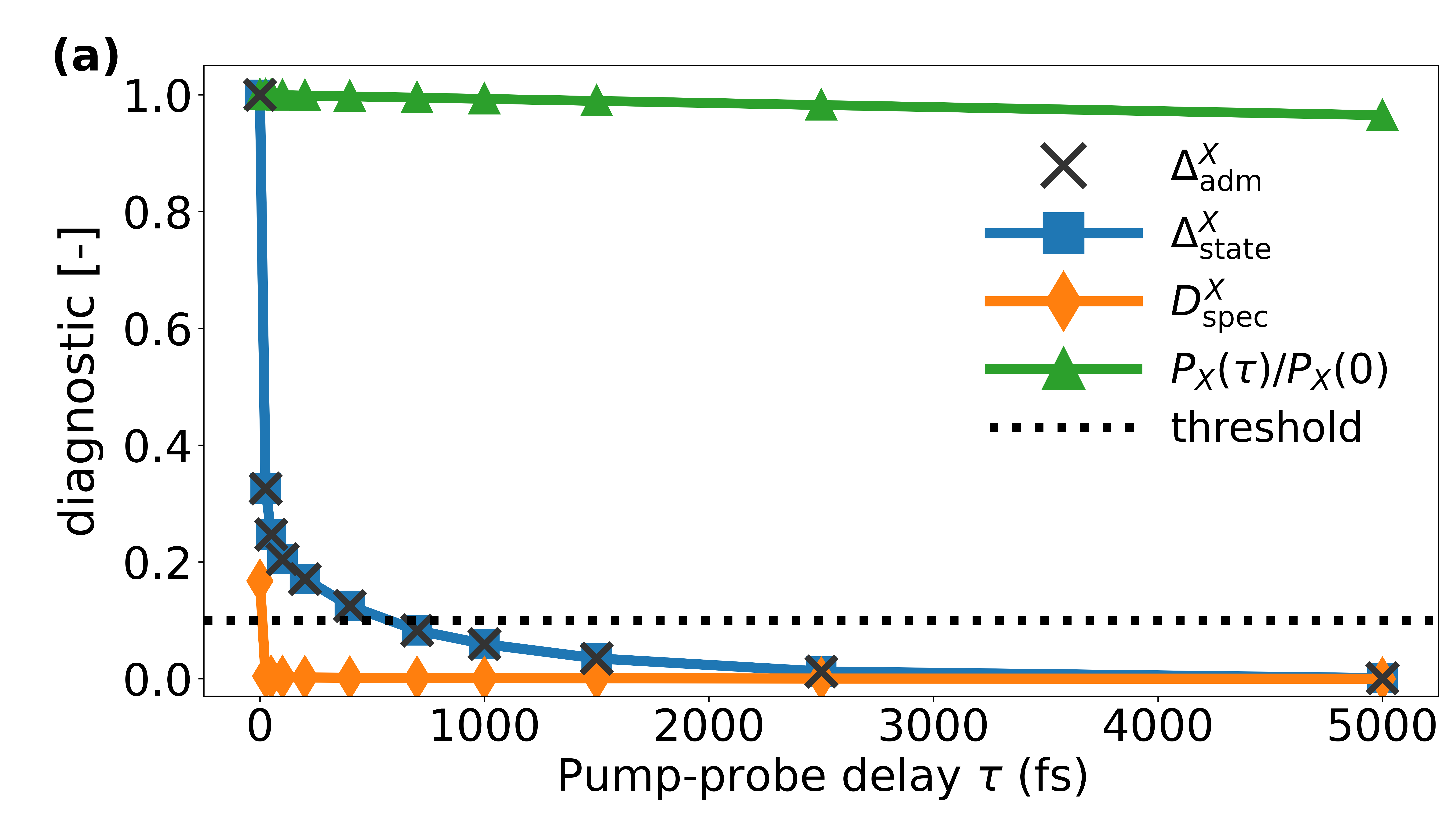}
  \vspace{0.35em}
  \maybeincludegraphics[width=\columnwidth]{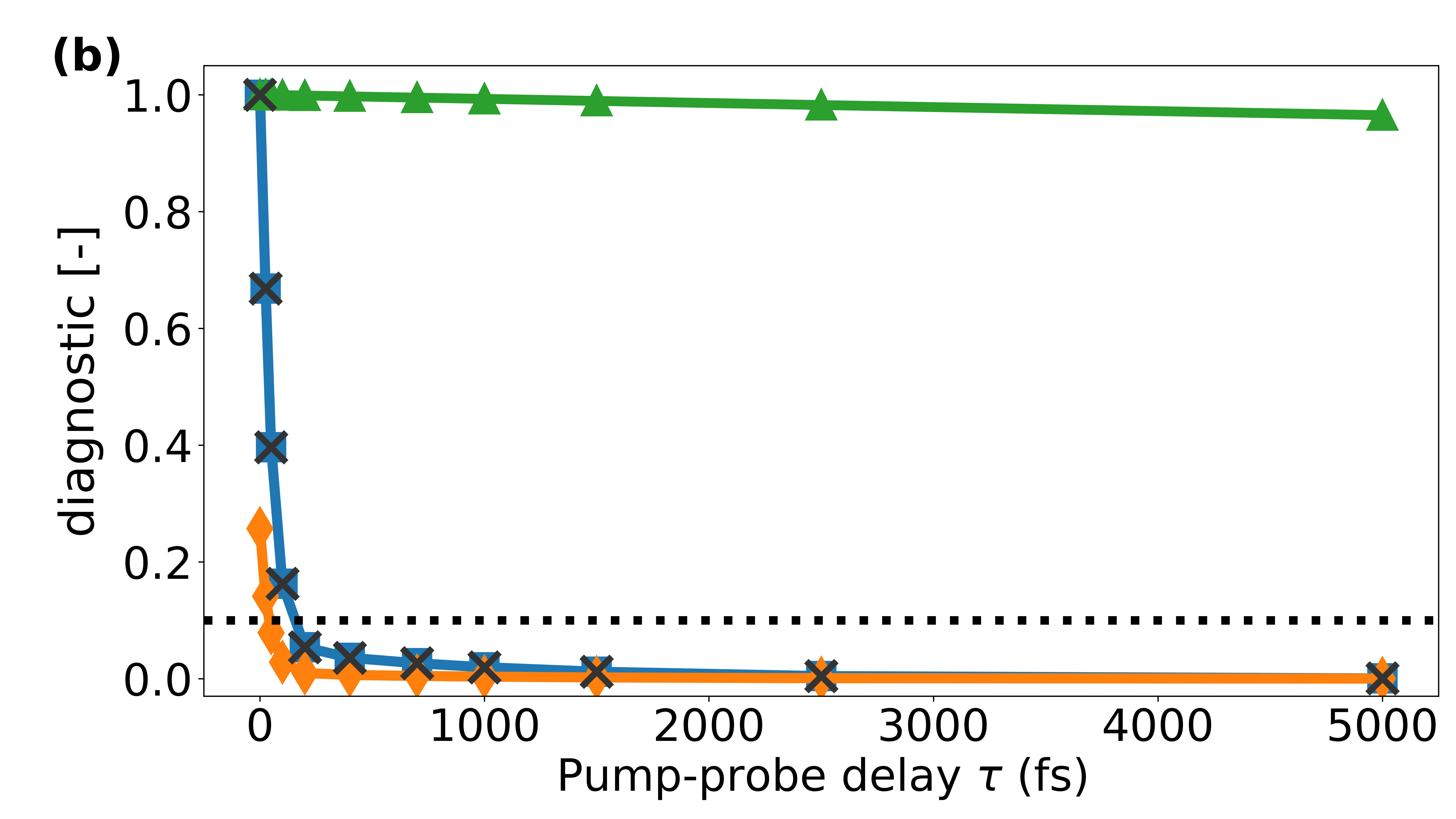}
  \vspace{0.35em}
  \maybeincludegraphics[width=\columnwidth]{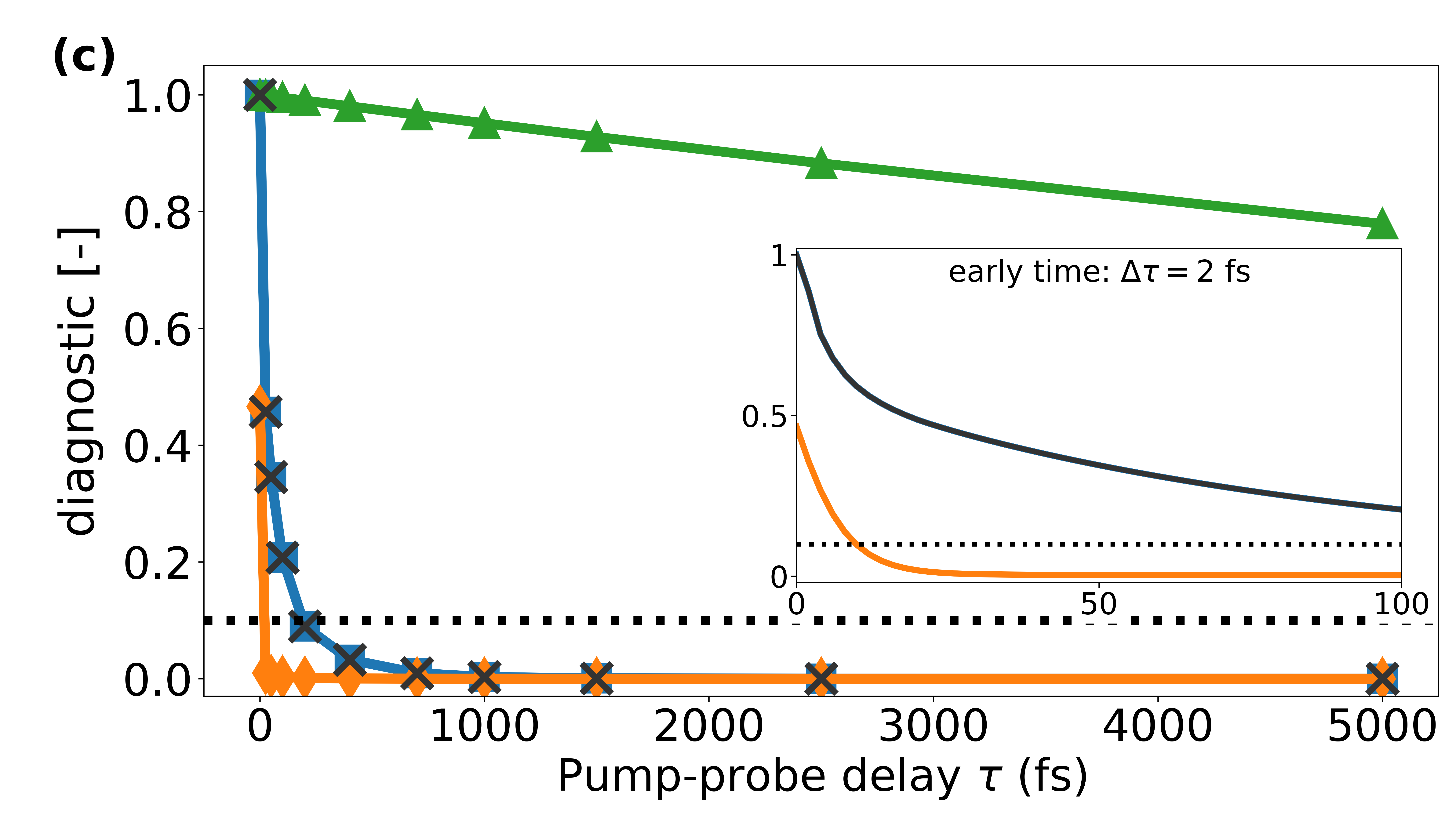}
  \caption{
  Delay-dependent conditional-exciton diagnostics for
  (a) the Ress squaraine-polymer squeezed helix,
  (b) the cisoid indolenine squaraine B hexamer, and
  (c) the helical perylene-bisimide stack.
  Blue squares and orange diamonds show
  \(\Delta_{\rm state}^{X}\) and \(D_{\rm spec}^{X}\), respectively.
  Black \(x\) symbols show the conservative admissibility envelope defined in
  Eq.~\eqref{eq:delta_adm_x}, and green triangles show the normalized
  one-exciton survival probability \(P_X(\tau)/P_X(0)\), which does not enter
  the envelope.
  The black dotted line marks the operational threshold of \(0.1\).
  The inset in panel (c) shows a separately evaluated benchmark-C calculation
  from \(0\) to \(100~{\rm fs}\) with \(\Delta\tau=2~{\rm fs}\); it is not an
  interpolation of the coarse-grid results.
  }
  \label{fig:organic_diagnostics}
\end{figure}

In the one-exciton Frenkel basis, the field-free exciton Hamiltonian is:
\begin{equation}
H_X
=
\sum_{i=1}^{N}
\varepsilon_i \ket{i}\bra{i}
+
\sum_{i<j}
J_{ij}
\left(
\ket{i}\bra{j}
+
\ket{j}\bra{i}
\right),
\label{eq:HX_benchmark}
\end{equation}
where \(\ket{i}\) denotes a local excitation on the chromophore \(i\), \(\varepsilon_i\) is the site energy, and \(J_{ij}\) is the excitonic coupling. The electronic ground state is denoted by \(\ket{g}\). The reduced E1--M1 response is generated by the effective transition operators [Eq.~\eqref{eq:effective_mu_m}], which are used to compare the pump-prepared retarded TRCD-like response with the conditional exciton reference. 

For benchmark A, we reconstructed the structure from the monomer coordinates and charges reported in Ref.~\citenum{Ress2023_ChemSci}, replicated into \(N=16\) chromophores on a left-handed helical scaffold. 
In this system, the transition-charge coupling (in eV) is evaluated as:
\begin{equation}
J_{mn}^{\rm TrEsp}
=
14.399645
\sum_{I\in m}
\sum_{J\in n}
\frac{
q_I^{\rm tr}q_J^{\rm tr}
}{
R_{IJ}/\text{\AA}
},
\label{eq:J_tresp}
\end{equation}
where \(q_I^{\rm tr}\) is the transition charge in units of \(e\), \(R_{IJ}\) is the distance in \AA{} between charge sites \(I\) and \(J\), and the numerical prefactor accounts for the conversion to eV \AA/e$^2$. The electrostatic site-energy correction (again in eV) is computed as
\begin{equation}
\delta \varepsilon_m
=
14.399645
\sum_{n\neq m}
\sum_{I\in m}
\sum_{J\in n}
\frac{
\left(q_I^{\rm exc}-q_I^{\rm gs}\right)q_J^{\rm gs}
}{
R_{IJ}/\text{\AA}
}.
\label{eq:site_shift_cdc}
\end{equation}
The site energy is then
\(\varepsilon_m=E_0+\delta\varepsilon_m\).
For the squeezed-helix benchmark, the reference transition wavelength is
\(708~{\rm nm}\), corresponding to
\(E_0=hc/(708~{\rm nm})=1.7514~{\rm eV}\).
The quantity \(J_{mn}^{\rm TrEsp}\) in Eq.~\eqref{eq:J_tresp} is the raw
transition-charge coupling. The coupling used in the Frenkel Hamiltonian is
\(J_{mn}=f_{\rm exc}J_{mn}^{\rm TrEsp}\), with \(f_{\rm exc}=2.4\), following
the Ress-type squeezed-helix parametrization~\cite{Ress2023_ChemSci}.

\begin{figure*}[!htbp]
  \centering
  \maybeincludegraphics[
    width=\textwidth
  ]{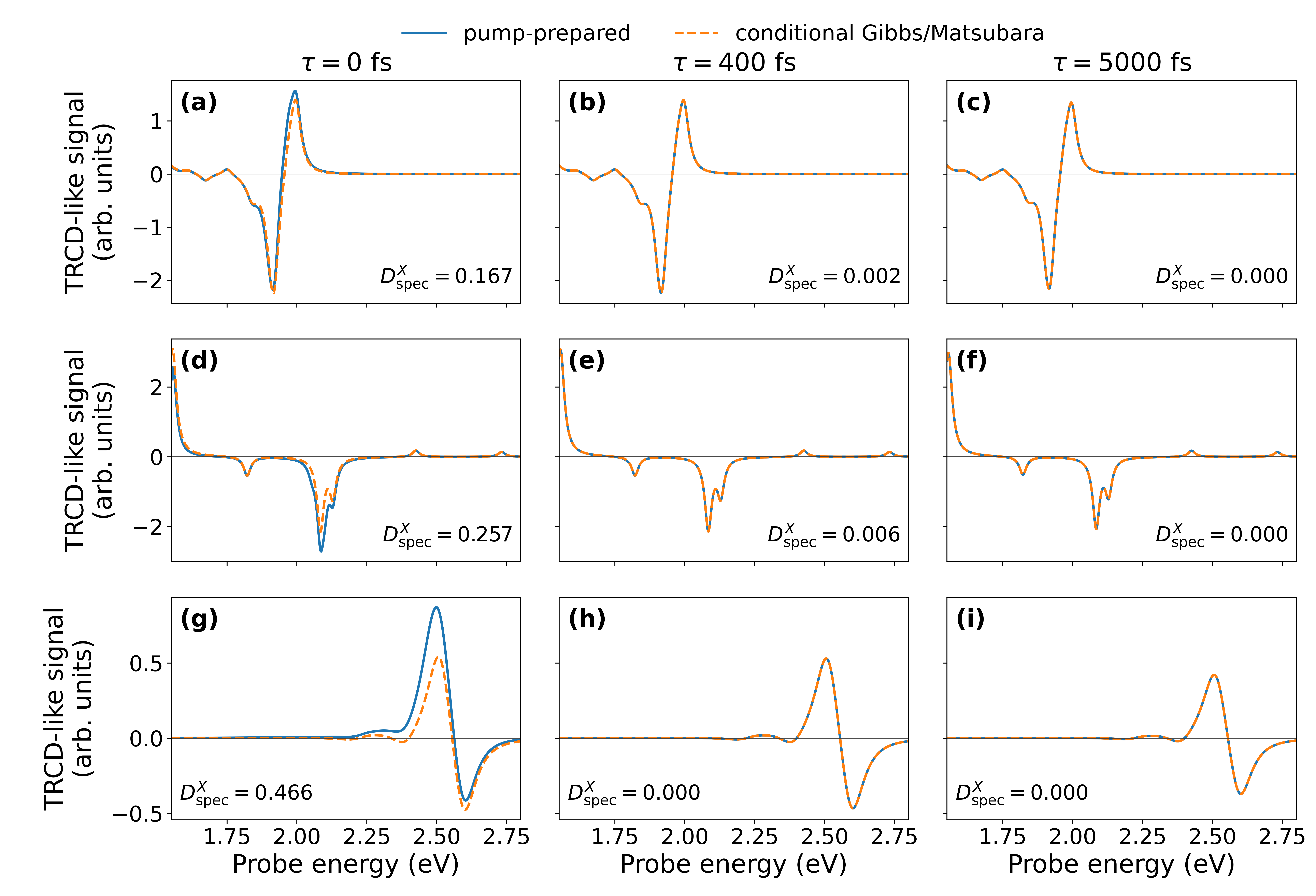}
\caption{
TRCD-like spectra for the three organic benchmarks.
Rows correspond to the Ress squaraine-polymer squeezed helix, the cisoid
indolenine squaraine B hexamer, and the helical perylene-bisimide stack.
Columns correspond to \(\tau=0\), \(400\), and \(5000~{\rm fs}\).
Solid curves show the pump-prepared responses and dashed curves show the
conditional Gibbs references.
The spectra are evaluated using the reciprocal mixed response of
Eq.~\eqref{eq:reduced_trcd_sum} and the E1--M1 pump preparation of
Eqs.~\eqref{eq:pump_weight_e1m1} and
\eqref{eq:pump_bright_initialization}.
The value printed in each panel is the integrated absolute spectral distance
\(D_{\rm spec}^{X}\) defined in Eq.~\eqref{eq:d_spec_x}.
Early-delay differences reflect the nonthermal pump-prepared distribution,
whereas late-delay overlap identifies the regime in which the conditional
stationary reference reproduces the selected E1--M1 response channel.
}
  \label{fig:organic_spectra}
\end{figure*}

For benchmark B, chosen to test the admissibility criterion in a small organic aggregate where structural disorder, nonuniform couplings, and exciton localization  are physically relevant, the cisoid indolenine squaraine is treated as a finite \(N=6\) aggregate, using the SQB monomer transition-energy scale and a nonuniform Coulomb-coupling pattern representative of the helical conformation adopted in acetone solution. This replaces a generic nearest-neighbour chain with a finite coupling matrix motivated by the SQB hexamer exciton-model calculations. 

For benchmark C, the PBI stack is kept as a literature-constrained Frenkel model~\cite{Sung2015_NatCommun}, with oligomeric helical stacks containing at least ten monomers, a helical angle of approximately \(30^\circ\), excitonic couplings on the order of \(700~{\rm cm^{-1}}\), and ultrafast coherent Frenkel-exciton dynamics. In the absence of a transferable PDB/transition-charge matrix, we simulated benchmark C as an extended helical \(\pi\)-stack.

In all benchmarks, the pump-prepared exciton state is initialized as a
nonthermal distribution in the exciton eigenbasis. To remain consistent with
the E1--M1 pump Hamiltonian in Eq.~\eqref{eq:pump_interaction}, the reduced excitation weight of exciton state \(\alpha\) for circular-pump
helicity \(h=\pm1\) is
\begin{align}
W_\alpha^{(h)}
&=
\Big[
F_\alpha^{\rm E1}
+
h\,s_{\rm M1}
\left(
\mathcal R_\alpha^{m\mu}
+
\mathcal R_\alpha^{\mu m}
\right)
\Big]_{+}
\nonumber\\
&\quad\times
\exp\!\left[
-\frac{
(E_\alpha-\hbar\omega_{\rm pump})^2
}{
2\sigma_{\rm pump}^2
}
\right],
\label{eq:pump_weight_e1m1}
\end{align}
where
\(F_\alpha^{\rm E1}=|\boldsymbol{\mu}_{\alpha g}|^2\), with
\(\boldsymbol{\mu}_{\alpha g}
=
\sum_n C_{n\alpha}\boldsymbol{\mu}_n\),
is the reduced electric-dipole excitation strength,
\(s_{\rm M1}\) is the reduced magnetic-interference scale, and
\([x]_{+}=\max(x,0)\) guarantees a non-negative excitation probability.
The
initial population is then
\begin{equation}
P_\alpha(0)
=
P_X(0)
\frac{
W_\alpha^{(h)}
}{
\sum_\nu W_\nu^{(h)}
}.
\label{eq:pump_bright_initialization}
\end{equation}
The calculations reported here use \(h=+1\) and \(s_{\rm M1}=1\). For the
electric-dipole-only control calculation, \(h=0\) is used as a numerical switch
that suppresses the helicity-sensitive interference term; it does not
represent a circular-pump helicity.
Terms quadratic in the magnetic interaction
are omitted consistently with the E1--M1 truncation used for the TRCD-like
response.

The populations are propagated with detailed-balance intramanifold rates,
slow one-exciton loss, and the coherence proxy in
Eq.~\eqref{eq:coherence_proxy_benchmarks}. The benchmark implementation does
not propagate an independent matrix of off-diagonal coherences.

At each pump--probe delay \(\tau\), we compute the state-level defect
\(\Delta_{\rm state}^{X}\), the spectral defect \(D_{\rm spec}^{X}\), and the
conservative admissibility envelope defined in Eq.~\eqref{eq:delta_adm_x}.

The threshold \(\Delta_{\rm adm}^{X}=0.1\) is used as an operational tolerance
rather than as a universal thermodynamic boundary. Because \(D_{\rm spec}^{X}\)
is normalized by the integrated absolute intensity of the pump-prepared
TRCD-like signal, \(D_{\rm spec}^{X}=0.1\) corresponds to a 10\% integrated
spectral deviation for the selected observable. We apply the same numerical
tolerance to the state-level defect to keep the screening criterion
conservative. In applications with higher experimental noise, broader spectral
resolution, or qualitative objectives, a looser tolerance may be sufficient,
whereas quantitative line-shape reconstruction may require a stricter value.

In practical calculations where the explicit pump-prepared TRCD-like spectrum
is not available, the initial admissibility decision can be made from the
state-level gate alone. The required inputs are the exciton Hamiltonian, the
pump-prepared one-exciton density matrix, and the relaxation/dephasing model
used to propagate it. The state-level gate is passed when the normalized
exciton populations are sufficiently close to the conditional Gibbs
distribution and the remaining coherence contribution is below the chosen
tolerance. In the reduced benchmarks, the latter contribution is represented
by the coherence-memory proxy of
Eq.~\eqref{eq:coherence_proxy_benchmarks}.

The survival probability \(P_X(\tau)/P_X(0)\) is monitored separately. It does
not determine mathematical admissibility, but indicates whether sufficient
excited-state population remains for the transient signal to be experimentally
meaningful. The spectral distance \(D_{\rm spec}^{X}\) is likewise not required
for the initial state-based screening; it is used here as an a posteriori
validation that the selected TRCD-like response is actually reproduced by the
conditional reference.

\section{Results and discussion}
\label{sec:results_discussion}

\begin{figure}[!htbp]
  \centering
  \maybeincludegraphics[width=\columnwidth]{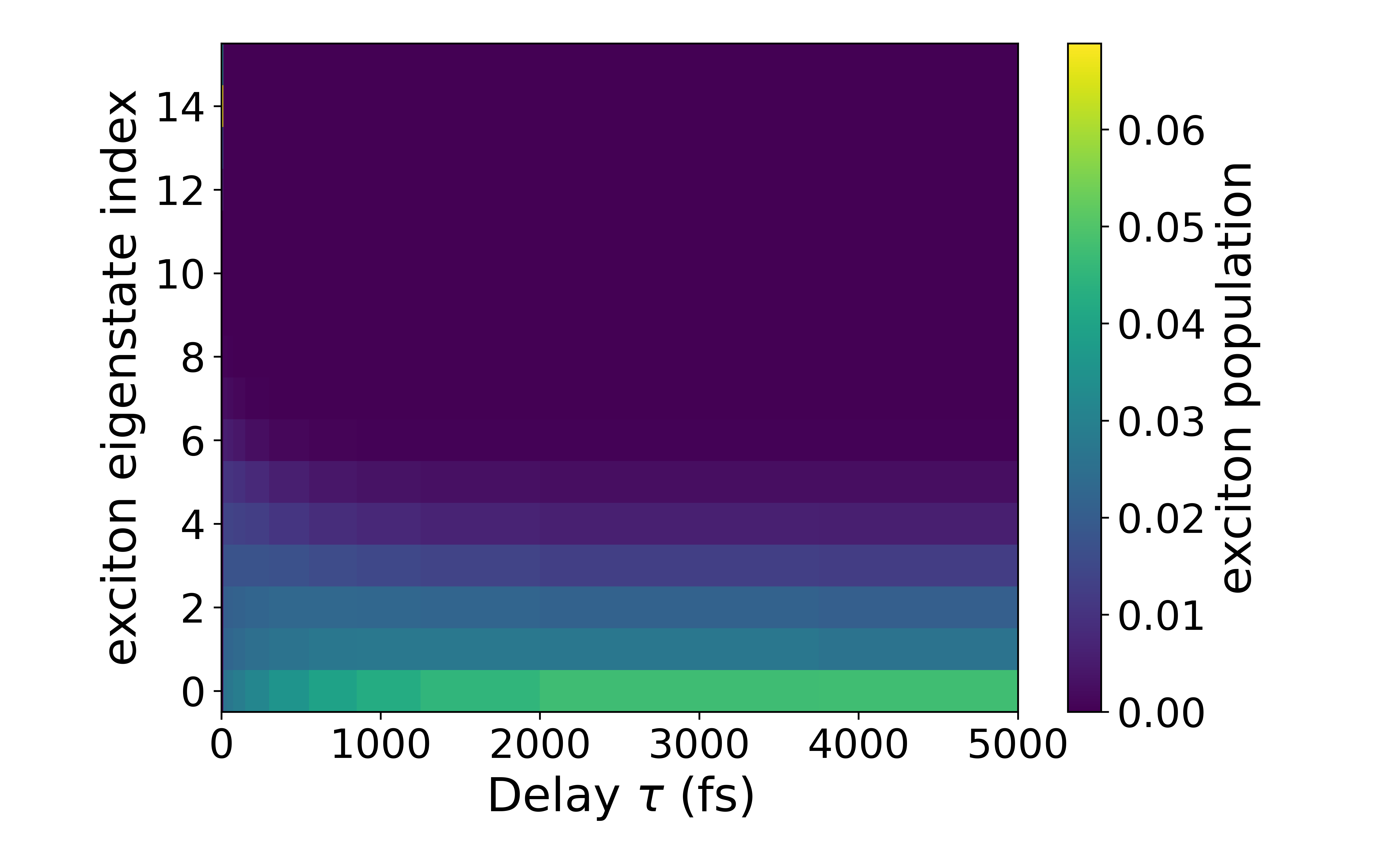}
  \vspace{0.35em}
  \maybeincludegraphics[width=\columnwidth]{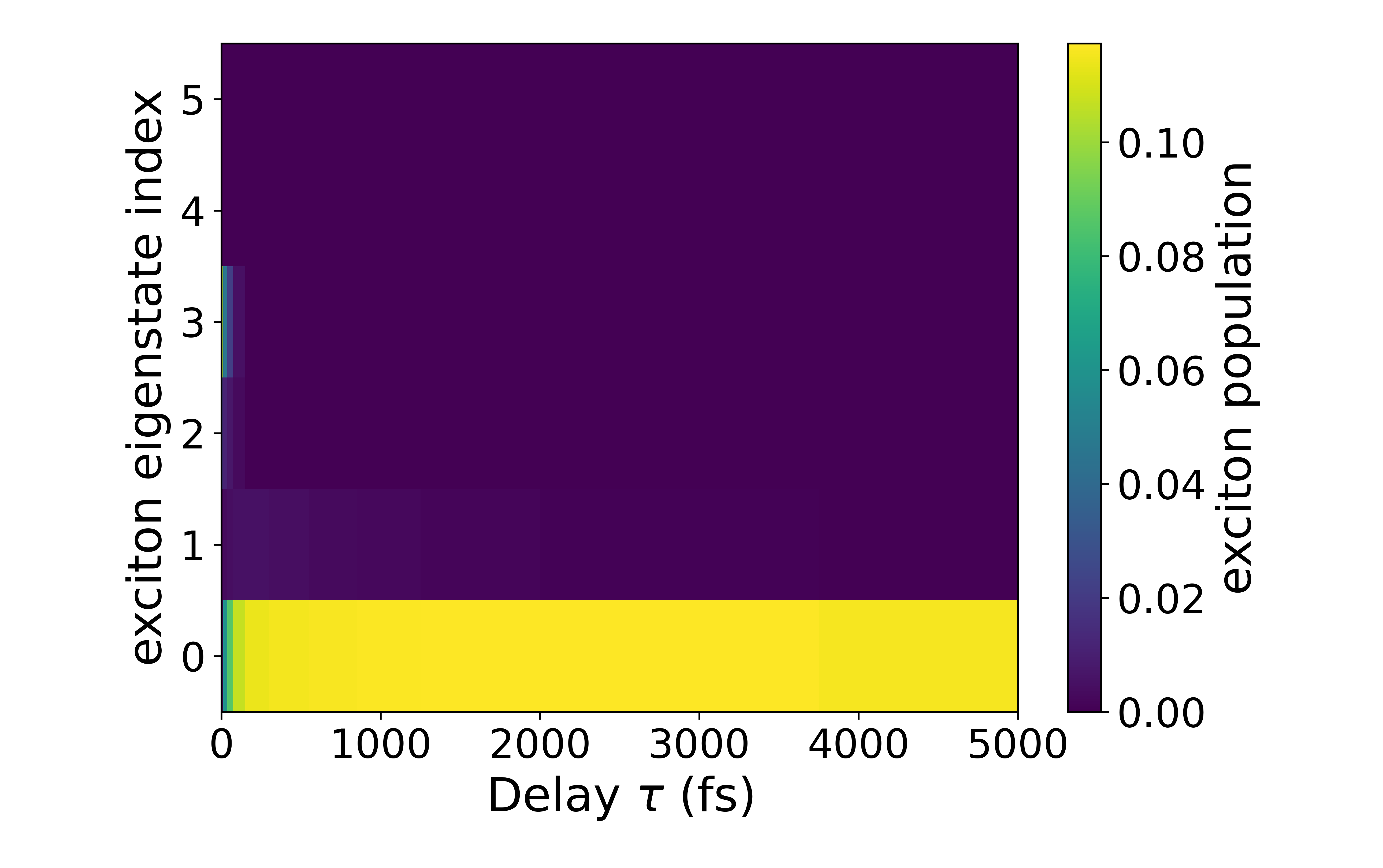}
  \vspace{0.35em}
  \maybeincludegraphics[width=\columnwidth]{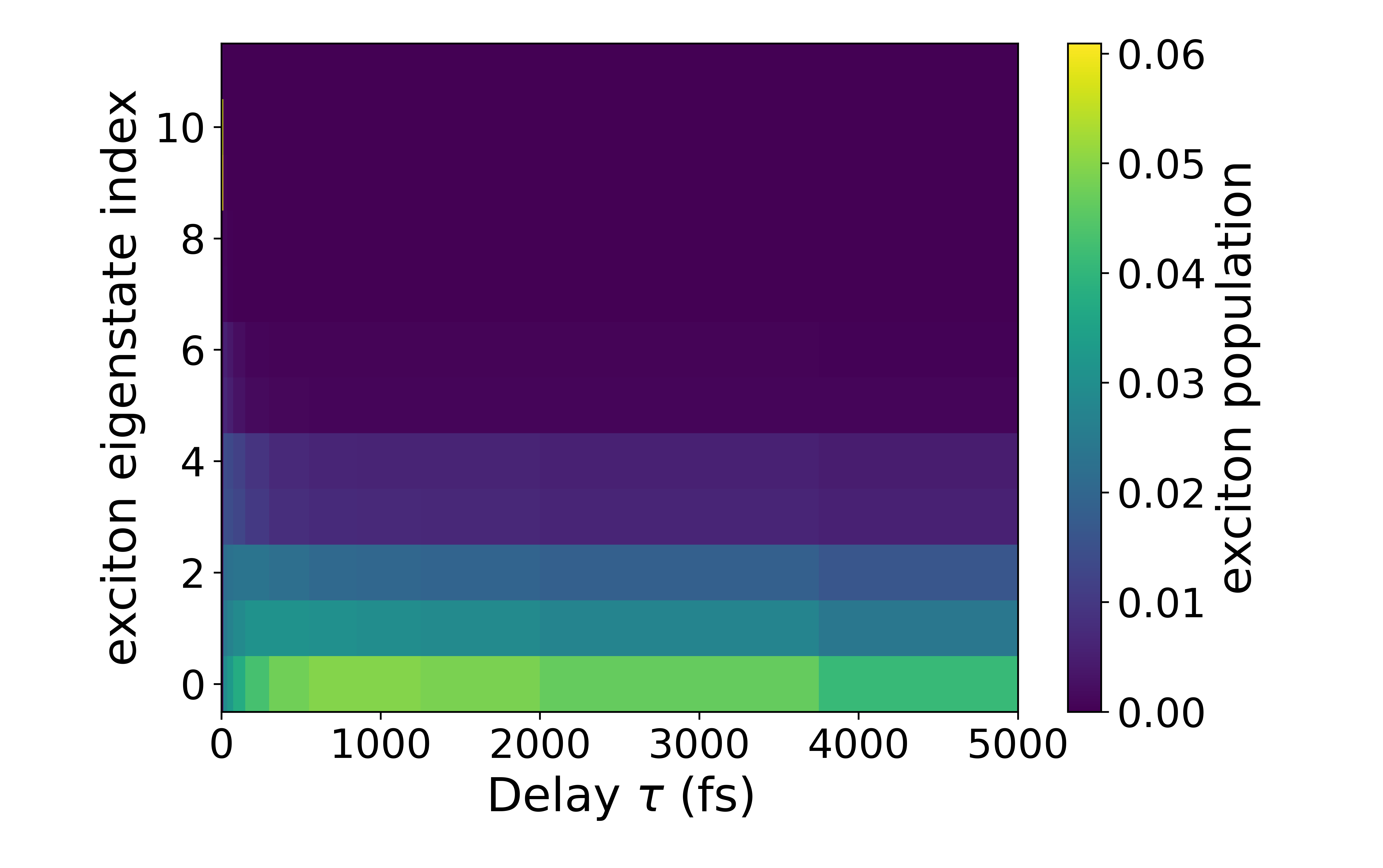}
\caption{
Exciton-eigenstate population maps for the three organic benchmarks:
top, Ress squaraine-polymer squeezed helix;
middle, cisoid indolenine squaraine B hexamer;
bottom, helical perylene-bisimide stack.
In each map, the horizontal axis is the pump--probe delay and the vertical axis is the exciton eigenstate index.
The color scale gives the population of each exciton eigenstate.
The maps show redistribution from the initially prepared nonthermal bright-state distribution toward the lower-energy part of the one-exciton manifold.
This intramanifold relaxation reduces the state-level contribution \(\Delta_{\rm state}^{X}\) entering the admissibility envelope \(\Delta_{\rm adm}^{X}\).
}
  \label{fig:organic_populations}
\end{figure}

Figure~\ref{fig:organic_diagnostics} summarizes the delay-dependent
diagnostics for the three benchmarks. The conservative envelope defined in
Eq.~\eqref{eq:delta_adm_x} opens the admissibility gate only when both the
state-level mismatch and the response-level spectral distance are below the
chosen tolerance. It therefore separates the genuinely nonequilibrium
early-time regime from the delays at which the conditional stationary
reference reproduces the selected TRCD-like response.

The normalized one-exciton population \(P_X(\tau)/P_X(0)\) provides
complementary information. It does not enter the admissibility defect. Instead,
it monitors whether the system still carries enough excited-state population for
a TRCD-like signal to be meaningful. The important point is that the
admissibility gate can open before \(P_X\) has decayed to zero, showing that
internal stationarity within the one-exciton manifold can be reached before
complete excited-state recovery.

Benchmark A, the reconstructed Ress-type squaraine-polymer squeezed helix,
shows a gradual decrease of the admissibility defect. The pump-prepared state
is initially non-admissible, and both its intramanifold distribution and its
TRCD-like spectrum differ from the conditional reference. The spectral
distance becomes small before the state-level defect crosses the threshold,
so the opening of the gate is controlled primarily by
\(\Delta_{\rm state}^{X}\). The one-exciton population remains close to its
initial value over this relaxation window, consistent with the separation
between intramanifold relaxation and slower population recovery reported for
the Ress-type aggregate~\cite{Ress2023_ChemSci}.

Benchmark B, the cisoid indolenine squaraine B hexamer, exhibits faster
state-level relaxation. Its initial spectral distance is finite, but the
pump-prepared and conditional-reference spectra become nearly
indistinguishable by the intermediate delay. On the sampled delay grid,
\(\Delta_{\rm state}^{X}\) falls below the \(0.1\) threshold between
\(100\) and \(200~{\rm fs}\), while a substantial one-exciton population is
still present. This behavior is consistent with the finite, structurally
nonuniform SQB hexamer model and the exciton-localization picture discussed
for disordered cisoid squaraine aggregates~\cite{Fischermeier2024_PCCP}.

Benchmark C, the helical perylene-bisimide stack, also displays rapid spectral
relaxation, consistent with the ultrafast coherent Frenkel-exciton dynamics
observed in self-assembled PBI stacks~\cite{Sung2015_NatCommun}. The dense
early-time calculation gives
\(D_{\rm spec}^{X}=0.466\), \(0.097\), \(0.018\), and \(0.003\) at
\(\tau=0\), \(10\), \(20\), and \(100~{\rm fs}\), respectively. Thus, the
observable-level mismatch falls below the \(0.1\) tolerance within
approximately \(10~{\rm fs}\). The \(2~{\rm fs}\) sampling shown in
Fig.~\ref{fig:organic_diagnostics}(c) resolves this decrease as a rapid but
continuous crossover. The full gate opens later because
\(\Delta_{\rm state}^{X}\) remains the limiting contribution.

Figure~\ref{fig:organic_spectra} shows the corresponding reciprocal E1--M1
responses. At \(\tau=0\), replacing the pump-prepared state by the conditional
Gibbs reference produces visible changes in line shape, signed spectral
weight, and peak amplitudes. At \(\tau=400~{\rm fs}\), benchmarks B and C are
already inside or close to the admissible regime, whereas benchmark A retains
a finite state-level mismatch. By \(\tau=5000~{\rm fs}\), the
pump-prepared and conditional-reference spectra overlap within the benchmark
tolerance for all three systems.

The population maps in Fig.~\ref{fig:organic_populations} provide the
complementary microscopic picture. Intramanifold relaxation redistributes the
initially prepared nonthermal population toward the lower-energy exciton
states. This redistribution is gradual and distributed over several low-energy
states in benchmarks A and C, whereas benchmark B exhibits a more rapid
accumulation in the lowest exciton state. At the same time, the phenomenological
coherence memory decays according to
Eq.~\eqref{eq:coherence_proxy_benchmarks}. Together, population relaxation and
the loss of coherence memory reduce \(\Delta_{\rm state}^{X}\) without
requiring complete decay of the one-exciton population.

Taken together, the benchmarks show that admissibility is both system
dependent and observable dependent. The gate identifies the delay at which a
conditional stationary reference becomes adequate for the selected TRCD-like
channel while a measurable one-exciton population is still present.

\section{Summary and Conclusions}
\label{sec:conclusions}

We have introduced a KMS-gated reconstruction protocol for determining when
the transient chiroptical response of a molecular aggregate can be represented
by a conditional imaginary-time reference. Within the weak-probe regime, the
measured TRCD signal is a causal reciprocal E1--M1 response of the
pump-prepared ensemble. The proposed diagnostic combines a state-level
comparison with an observable-level spectral validation and therefore
distinguishes delays that require explicit real-time treatment from those at
which the selected response channel has become conditionally stationary.

Across the three literature-constrained benchmarks, early nonthermal
distributions remain non-admissible, whereas intramanifold population
relaxation and the decay of coherence memory eventually open an admissible
delay window. This stationarity is reached within the optically active
one-exciton manifold and does not require complete ground-state recovery.
Imaginary-time methods may therefore become applicable while a measurable
excited-state population is still present, provided that the selected
response has lost its sensitivity to the pump-prepared distribution.

The present implementation is most natural for aggregates where a one-exciton
manifold can be separated from the ground state and where the relevant nuclear
structure is either slowly varying or incorporated into an effective Frenkel
Hamiltonian. For smaller molecules with sparse excited states, strong vibronic
mixing, conical-intersection dynamics, or large-amplitude nuclear motion during
the pump--probe delay, a purely electronic conditional Gibbs reference may be
insufficient. In such cases, the same admissibility idea can still be applied,
but the reference ensemble must be enlarged to include vibronic or nuclear
degrees of freedom, or the response must be propagated explicitly in real time.

Although demonstrated here with parameterized Frenkel-exciton models, the
formalism is general and can be extended to \textit{ab initio}
workflows~\cite{krum+24apr}. In particular, coupling this protocol with
gauge-invariant real-time time-dependent density functional
theory~\cite{krum+20jcp,trep+25jpcl} or Bethe-Salpeter-based exciton
dynamics~\cite{qiao2026first} will enable efficient screening of nonequilibrium
chiroptical spectra in complex systems by automatically switching between
real-time and imaginary-time many-body treatments.

\section*{Author contributions}
Christian Tantardini: Conceptualization (lead), Methodology (lead), Investigation (lead), Formal analysis (lead), Visualization (equal), Writing – Original Draft (lead), Writing – Review \& Editing (supporting).

Caterina Cocchi: Funding acquisition (lead), Validation (lead), Writing – Review \& Editing (lead).

\section*{Conflicts of interest}
There are no conflicts to declare.

\section*{Data availability}
The Python code, input files, benchmark parameters, and numerical outputs required to reproduce the three organic Frenkel-exciton benchmarks are publicly available at: \url{https://github.com/Christian48596/kms-trcd-exciton-benchmarks}. Using this repository, the admissibility diagnostics, TRCD-like spectra, and exciton-population maps reported in this work can be reproduced.

\section*{Acknowledgements}
C.C. acknowledges funding from the German Research Foundation (DFG), project numbers 524452181 (INPULS) and 398816777 (CRC 1375, subproject A08).

\bibliography{main}

\end{document}